\documentclass{aa}  
\usepackage{graphicx}
\usepackage{txfonts}
\usepackage{lipsum}
\usepackage{subcaption}         % necessary for continued figures, example in section 3
\usepackage[table]{xcolor}      % and appendix
\usepackage{lscape}             % to rotate a single page table, example in appendix.
\usepackage{placeins}           % useful with \FloatBarrier, to keep 
\usepackage[scr]{rsfso}
\newcommand{\Laplace}{\mathscr{L}}
\begin{document}

   \title{Towards Inter-satellite Laser Metrology for Formation Flying Interferometry}

%%%%%%%%%%%%%%%%%%%%%%%%%%%%%%%%%%%%%%%%
% Please do not include ORCIDs next to author names.
% Only ORCIDs authenticated by individual authors in EDP Sciences editorial system will be taken into account.
% ORCIDs included here will be removed.
%%%%%%%%%%%%%%%%%%%%%%%%%%%%%%%%%%%%%%%%

   \author{A.O. Chan \inst{1,2}\fnmsep\thanks{Corresponding author: anika.chan@anu.edu.au}
        \and J.E  Debs \inst{3}
        \and M.J. Ireland \inst{2}
        \and C.P. Bandutunga \inst{1}}

   \institute{Centre for Gravitational Astrophysics, College of Science and Medicine, Australian National University, Acton, Australian Capital Territory 2601, Australia
            \and Research School of Astronomy and Astrophysics, Australian National University, Canberra ACT 2601, Australia
            \and Department of Quantum Science and Technology, College of Science and Medicine, Australian National University, Acton, Australian Capital Territory 2601, Australia}

   %\date{Received September 30, 20XX}

% \abstract{}{}{}{}{}
% 5 {} token are mandatory
 
  \abstract
   {The direct detection and atmospheric characterization of terrestrial exoplanets has been highlighted as a key science focus by the US ASTRO2020 Decadal survey and the ESA Voyage 2050 plan. This will require a new generation of telescope which is able to meet the contrast requirements necessary to image a planet many orders of magnitude fainter than its host star - a technological feat which currently limits our observation capabilities. In the mid-infrared, the most viable solution has been shown to be a formation-flying nulling space interferometer.}
   {In this work, we present the initial prototype for an inter-spacecraft metrology system. This system is a sub-system of the Pyxis/SEIRIOS interferometer, a formation-flying interferometer which serves as a technology demonstrator for the LIFE mission.} 
   {We have designed a multi-stage metrology system using three interferometric measurement techniques, covering varying distance resolution ranges. Digital Interferometric (DI) ranging provides the coarsest measurement, while a differential phase readout provides sub-nm precision. Multi-wavelength interferometry is employed to overcome the phase ambiguity associated with the differential phase readout.}
   {A single-wavelength electronic noise floor below 20 pm RMS over 10 s was measured by applying a common photodetector signal to both readout channels, a factor of 28 below the 560 pm requirement. For the multi-wavelength readout, the obtained electronic noise RMS was 556\,nm at 10 second integration, decreasing to 200 nm below 3 second integration. The full optical measurement in laboratory air is currently limited by environmental path fluctuations.}
   {We demonstrate an initial prototype system for precision ranging between satellites for formation-flying nulling interferometers. }

   \keywords{Techniques: interferometric --
                Planets and satellites: terrestrial planets --
                Infrared: planetary systems
               }

   \maketitle

   \nolinenumbers

%%%%%%%%%%%%%%%%%%%%%%%%%%%%%%%%%%%%%%%%%%%%%%%%%%%%%%%%%%%%%%
\section{Introduction}
Since the first confirmed detection of an exoplanet in 1992 \citep{wolszczan1992planetary}, exoplanet research has become a major subfield of astrophysics, with over 7000 exoplanet detections made to date. In 2021, the ASTRO2020 Decadal Survey from the National Academies of Sciences, Engineering and Medicine recognized the field and identified the investigation of Earth-like exoplanets as a key science theme for the next decade \citep{monnier2019realistic}. ESA made a similar statement in the ESA Voyage 2050 plan \citep{snellen2019esa}. To fulfill the science themes identified in ASTRO2020, we must focus on obtaining spectra from the atmospheres of terrestrial exoplanets in order to learn about its constituents, its formation, and whether life could be present. However, detection and characterization of these planets is an ongoing challenge with current instrumentation, which has biased the population of detected planets on the basis of what is actually feasible to observe \citep{ortiz2025exoplanet}. This means that smaller Earth-like exoplanets orbiting Sun-like stars tend to go undetected, compared to larger high-SNR exoplanets orbiting dimmer stars \citep{scalo2007m}. Furthermore, transmission spectroscopy and direct detection are the only methods of exoplanet detection that can actually retrieve atmospheric spectra \citep{piaulet2024jwst} -- the former being geometrically limited \citep{wright2012exoplanet} and the latter being viable only to young large stars on the outer edges of their host solar system \citep{zurlo2024direct}. 

For direct imaging, terrestrial exoplanet observability is limited by the brightness contrast between the planet and its host star, which can range from $10^{-6}$ for a hot Jupiter to $\leq10^{-9}$ for an Earth-like exoplanet \citep{claudi2024high}. This ratio can be improved by observing in the infrared \citep{cagigal2001exoplanet}, which would require a space-based telescope to bypass the thermal background associated with a ground-based telescope. Extremely large ground-based telescopes (ELTs) \citep{zhang2024detecting} will also be powerful enough to resolve terrestrial exoplanets; however, they are limited to nearby stars only. In space, existing missions such as \textit{Kepler}, the Transiting Exoplanet Survey Satellite (TESS) and the \textit{James Webb} Space Telescope have already laid the groundwork for space-based exoplanet observations through their detections of several thousand planetary candidates. Unfortunately, none of these missions are capable of performing the detailed characterization of temperate exoplanet atmospheres required to meet the science goals outlined in ASTRO2020 and Voyage 2050.

Furthermore, a key bandwidth of interest for exoplanet spectroscopy is the mid-infrared (MIR) regime, since it encompasses known biosignatures that are not detectable or not present in the optical/near-UV wavelength regime \citep{quanz2018exoplanet}. In order to observe terrestrial planets in the MIR, contrast ratios on the order of $10^{-7}$ with respect to the host star must be overcome \citep{zhang2024detecting}. These high spatial resolution and sensitivity requirements can only be met through nulling, an interferometric technique that exploits the path length difference between the host star and planet to nullify the host star. 

Proposed by Bracewell in 1978 \citep{bracewell1978}, nulling interferometry involves multiple science telescopes at known relative distances observing a singular target. The telescopes are positioned such that when received light is interfered by a recombiner apparatus, an on-axis target is destructively interfered (nulled), while light from an off-axis source is constructively interfered. This allows us to position a target planet within a bright fringe, while light from the corresponding host star is nulled, effectively relaxing the contrast requirement. Additionally, a proposed MIR telescope system must be space-based, in order to bypass the thermal background associated with the Earth's atmosphere. In order to perform nulling interferometry in space, formation flying, ie. the controlled positioning of multiple satellites operating as a single constellation, is required. Specifically, a potential space-based interferometer would require formation flying technology capable of positioning multiple combiner satellites (science telescopes) to sub-wavelength displacements, relative to a recombiner satellite. No such mission has been launched as of this work. It is because of this, and the recommendations from ASTRO2020 and Voyage 2050, that the Large Interferometer for Exoplanets (LIFE) initiative was created.

The LIFE Collaboration \citep{glauser2024large} is a European-led mission aimed at launching the first MIR space-based formation flying nulling interferometer capable of probing thermal emission from terrestrial exoplanet atmospheres. LIFE aims to detect 30 to 50 terrestrial exoplanets, focusing on characteristics that are not possible to determine using transit or reflected light techniques, such as effective temperature and radius. It will also look for important molecular signatures such as water ($H_2O$), carbon dioxide ($CO_2$), ozone ($O_3$) and methane ($CH_4$) -- the presence of the latter two in exoplanetary atmospheres can currently only be explained by biological activity \citep{quanz2021atmospheric}. Technology for multiple subsystems is in development for LIFE, including the Nulling Interferometric Cryogenic Experiment (NICE) \citep{ranganathan2022nulling}, waveguides and integrated optics \citep{montesinos2024low}, detectors \citep{atkinson2020geosnap}, piezo-electric deformable mirrors \citep{huisman2021high}, and a compact delay line for space interferometry \citep{glauser2024large}. Formation flying technology, specifically the metrology required to enable formation flying fringe tracking, has not yet been demonstrated. 

As such, we present an initial prototype for a digitally addressable metrology subsystem for fringe acquisition and tracking; the components of which are all commercially available off the shelf and space viable. We are able to extract a sub-wavelength distance measurement by encoding the outgoing metrology laser with a phase shifting key, enabling real time phase tracking. This work functions as a subsystem to the SEIRIOS mission \citep{ikari2021seirios}, a Japanese-led formation flying fringe tracking science demonstrator for LIFE.

In Section \ref{sec:background}, we discuss the system requirements and provide background on the metrological techniques utilized in the system. Section \ref{sec:methods} outlines the optical and digital design. Static and dynamic measurements probing system stability and sensitivity are presented in Section \ref{sec:results}, as well as an analysis of phase-induced laser frequency noise. We summarize our findings and further work in Section \ref{sec:conclusion}.

%%%%%%%%%%%%%%%%%%%%%%%%%%%%%%%%%%%%%%%%%%%%%%%%%%%%%%%%%%%%%%
\section{Background}
\label{sec:background}
\subsection{Fringe Tracking and Acquisition}
\label{sec:fringetracking}
Interferometric measurements at infrared wavelengths are dependent on two key processes: fringe acquisition and fringe tracking. Fringe acquisition is the process of aligning two beams of light to produce an interferometric pattern. In order to find the correct alignment position, a piston actuator scans the delay line within a search window, in steps equal to the coherence length of the astronomical spectrograph. The step size $L_c$ is given by:
\begin{equation}
    L_c = \frac{\lambda^2}{\Delta \lambda}
\end{equation}
To detect fringes with a high SNR (ie. an SNR of 5 or higher), each step must be held long enough to detect a sufficient number of photons, particularly when observing fainter targets. The required total search length is dictated by the uncertainty in the relative positions of the spacecraft. This makes observing faint targets a complex control problem, which may require a significant amount of time on-orbit. For this reason, we have an absolute metrology requirement that the relative position uncertainty in the optical axis direction must be less than the longest coherence length of the beam combiners -- which is approximately 100 microns. This ensures that the fringes fall within the search window, significantly reducing fringe acquisition time. 

Any movement of the fringes during tracking will also reduce the SNR of the fringes during all individual exposures. Therefore, there is also a total phase excursion requirement of $\sim$1 radian of phase per exposure time, which applies in both the optical axis direction, and the direction perpendicular to the optical axis. 

\subsection{Inter-spacecraft Metrology System Requirements}
\label{sec:requirements}
\begin{figure*}[ht!]
   \centering
   \includegraphics[width=0.8\textwidth]{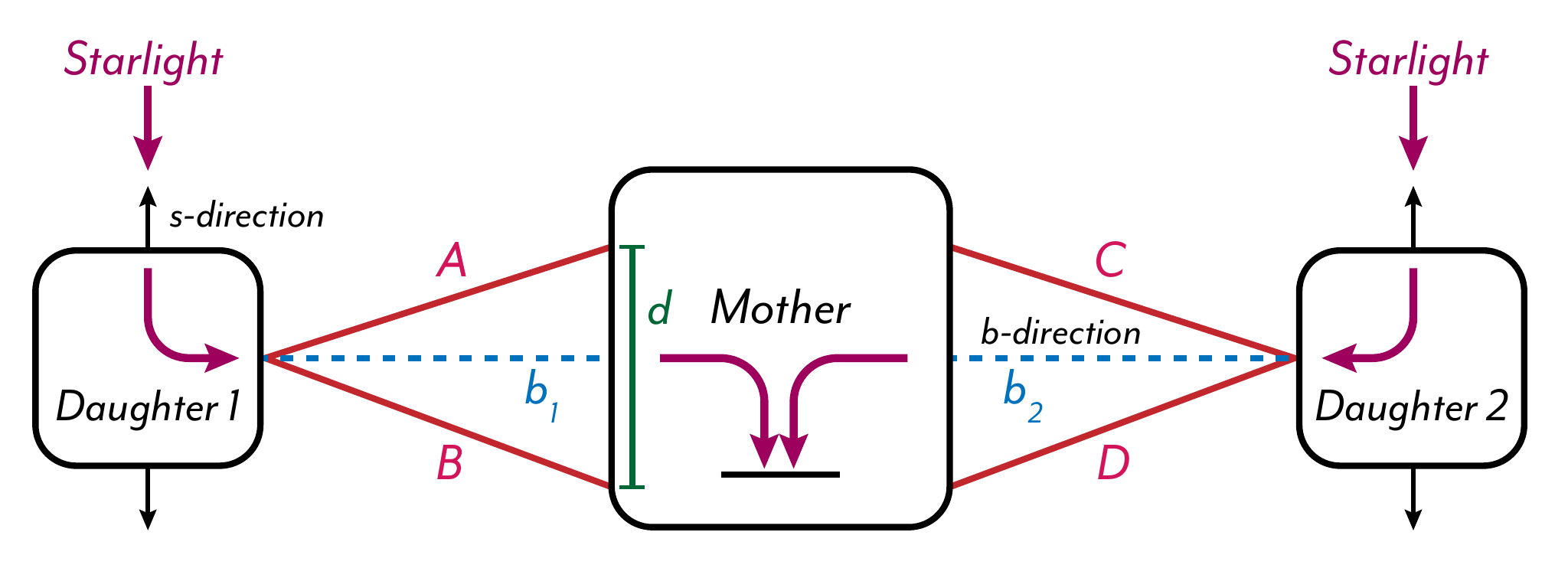}
      \caption{Schematic of the inter-spacecraft metrology system for Pyxis (or SEIRIOS), where s ('across') is the vertical positioning, and b ('along') is the horizontal or baseline positioning. $b_1$ and $b_2$ represent the absolute baseline distances between the daughters and the mother. d is the distance between the collimators on each side of the mother. With knowledge of the $b$ and $s$ direction offsets for each daughter satellite, the starlight path length can been matched. To obtain $s$-direction measurement, interferometric measurements are made over path lengths A, B, C and D \citep{lagadec2020compact, matsuo2022high}.}
        \label{satellitearrangement}
   \end{figure*}
In this work, our aim is to construct a system capable of tracking the dynamic displacement of the daughter satellites with respect to the mother satellite. Both daughter satellites collect starlight, and relay it to the mother satellite, where it is delay matched and interfered.. The successful combination of starlight at the mother satellite is dependent on sub-wavelength precision knowledge of the distance between satellites.

We define the baseline positioning as the b-direction, and the perpendicular direction as the s-direction. We also assume that we have \textit{a priori} knowledge of the b-direction displacement at the level of 1\,cm or better (e.g. from differential GPS in low-earth orbit).

The physical separation between the mother satellite and a daughter satellite, $b_1$ and $b_2$, can range from 5\,m (minimum) to 50\,m (maximum). Since the only metrology-related optics on the daughters are retroreflectors (to return metrological laser light), we must consider the roundtrip distance, ie. double the physical separation (10 to 100\,m). Each daughter has two laser interferometers to the mother: A and B for daughter 1, and C and D for daughter 2. Knowledge of the path difference between the laser interferometers for each daughter ($A-B$ and $C-D$) is essential to determine how close to zero the s-distance is with respect to the mother satellite. The system's ability to track distance is directly dependent on this s-distance parameter. Finally, the relative velocities between the satellites must be known to maintain positional stability over time, enabling the extraction of interferometric fringes with the highest possible contrast. All of the sensitivity requirements for this metrology system are listed in Table \ref{tab:requirements}. The path difference knowledge for s-direction, computed from $(A-B)$ or $(C-D)$ is one of the primary goals as it will enable the spacecraft to actively acquire starlight fringes without extended search ranges. The b-direction measurements $(A-C)$ or $(B-D)$ are a secondary priority as auxiliary sensors on the satellite bus will be able to provide positioning information that meets those requirements. 

\begin{table}[h!]
\centering
\begin{tabular}{|p{4cm}|p{3cm}|}
\hline
\textbf{Parameter} & \textbf{Requirement} \\
\hline
Minimum distance ($b_1$ \& $b_2$) & 5\,m (10\,m roundtrip) \\
\hline
Maximum distance ($b_1$ \& $b_2$) & 50\,m (100\,m roundtrip) \\
\hline
Collimator separation $d$ & 0.2\,m \\
\hline
Maximum s-displacement & 0.05\,m \\
\hline
Path difference knowledge (A - C) or (B - D) & 100\,\textmu m \\
\hline
Path difference knowledge (A - B) or (C - D) & 400\,nm\\
\hline
Total displacement sensitivity ($s_1 - s_2 + b_1-b_2$) & 100\,nm \\
\hline
Displacement error budget $s_1$ or $s_2$ & 70\,nm \\
\hline
Displacement error budget $b_1$ or $b_2$ & 10\,nm \\
\hline
Interferometer displacement sensitivity $L$  & 560\,pm RMS over 10 seconds \\
\hline
Interferometer velocity sensitivity  & 56\,pm/s \\
\hline
\end{tabular}
\caption{Summary of interferometer sensitivity and knowledge requirements for this metrology system.}
\label{tab:requirements}
\end{table}

\subsection{A Three Stage Metrology System}
\label{sec:threestage}
Interferometric measurements at the sensitivities required for this metrology system are arguably not that challenging. The technological challenge for this system is achieving this precision over the ranges outlined in Table \ref{tab:requirements}, a feat which is near impossible for a singular system. Therefore, we propose a multi-stage metrology system which synergizes multiple metrological techniques to cover increasingly finer measurement regimes. 

Starting with the coarsest measurement scheme, we encode all beams in this metrology system  with a time-dependent pseudo-random phase shifting sequence. Decoding of the returning signals is performed using a local copy of the phase shifting sequence. The time delay at which the returning signal and stored copy of the phase shifting sequence exhibit the highest auto-correlation properties is then inferred to be the time of flight of the signal, which is used to determine the distance traveled. Therefore, the distance resolution of this technique is dependent symbol rate of the phase shifting sequence. More explicitly, the distance resolution of this technique is determined by the distance a signal of interest could travel in the time it takes for an element in the code to change (chip rate). This measurement technique is known as Digital Interferometric (DI) ranging \cite{sutton2012digitally}. There are two major benefits to DI over conventional ranging techniques, one being that multiple signals can be identified in digital signal processing without requiring multiple photodiodes. Since it relies on the phase of the signal, DI can be also used to multiplex signals with the same time-of-flight, as long as their wavelengths are different. The second benefit is that since decoding of the signal of interest is both time-dependent and code-dependent, any signal that is not encoded with the phase shifting sequence (ie. noise) will be convolved with the local copy of the code during the decoding process, pushing it into a higher frequency band. This noise can then be filtered out using a digital low pass filter, improving the SNR of the measurement. 

To obtain sub-wavelength displacement knowledge, we extend upon DI by utilizing in-phase (I) and quadrature (Q) demodulation to obtain a phase measurement of the returning signal. The pseudo-random sequence mentioned previously is specifically a four level quadrature phase shifting key (QPSK) code, which when encoded onto incoming laser light, scrambles the phase of the light by shifting pseudo-randomly with four different phase values. The QPSK code is generated using two bipolar maximal length sequences (m-sequences), a type of binary pseudo-random sequence, following the relation: 
\begin{align}
    R(t) = \frac{\pi}{2}(x_2(t)-\frac{1}{2}x_2(t)x_1(t))
\end{align}
where $x_1$ and $x_2$ are the m-sequences comprising of values $\pm1$. When the QPSK code is encoded onto the laser light, it shifts the phase of the light, at the chip rate, by one of four possible QPSK values: $\pm\frac{3\pi}{4}$ or $\pm\frac{\pi}{4}$. This evenly samples the four quadrants of the interferometric fringe, generating a phase constellation that is used as a reference to unscramble the returning signal. Upon decoding the returning signal, the phase offset between the initial phase-shifted light, and the decoded return signal tells us the phase associated with the distance traversed by the light.

The phase readout determines the round trip path difference $L$ by modulo the optical wavelength, $\lambda$. The integer fringe order is lost, and therefore the range is ambiguous beyond the wavelength. During continuous operation this is not a limitation, as the phase can be unwrapped back to track changes in pathlength. The issue arises when the link is initialized, either on startup or after an outage. In this instance, a distance baseline must be reestablished to reacquire fringes on starlight which requires a length knowledge of 400 nm on the difference between the ranges A and B in Fig. \ref{satellitearrangement}. This is beyond the range gating capabilities of DI, which is on the metre scale. We therefore add an intermediate stage based on multi-wavelength interferometry (MWI) \cite{deGroot1994twocolour,amann2001laser-4ca, mcsorley2025digital-1f7}.

This technique involves taking phase measurements at two different wavelengths, where the difference produces a synthetic beat frequency $\nu_{beat}$. Two signals at different frequencies ($\nu_1$ and $\nu_2$) are encoded using a uniquely delayed $R(t)$. The field at the photodetector under multi-wavelength operation is then given by:
\begin{align}
    E_{\text{PD}} &= A_M\exp(-i[\omega_1t + R(t - \tau_{M}-\tau_1)]) \nonumber\\
    &\quad+ A_D\exp(-i[\omega_1t + \Delta\phi_1(t) + R(t - \tau_{D}-\tau_1)])\nonumber\\
    &\quad+ A_M\exp(-i[\omega_2t + R(t - \tau_{M}-\tau_2)]) \nonumber\\
    &\quad+ A_D\exp(-i[\omega_2t + \Delta\phi_2(t) + R(t - \tau_{D}-\tau_2)]),
\end{align}
which consists of four fields, two at each wavelength $\omega_1$ and $\omega_2$, and two from the mother ($M$) and daughter ($D$) reflections that comprise each interferometer. The delays $\tau_D$ and $\tau_M$ are the physical time-of-flight delays to the daughter and mother satellites respectively. The delays $\tau_1$ and $\tau_2$ are the digital delays used to identify wavelengths 1 and 2. This yields four unique codes at the detector corresponding to each field, which can be pairwise decoded. This allows each interferometer to individually decode to retrieve the phases, $\Delta\phi_1(t)$ and $\Delta\phi_2(t)$, associated with the distance measured by each wavelength between the mother and daughter satellites. 

The difference between these phases is equivalent to the phase of $\nu_{MW}$, where the corresponding beat wavelength can be many orders of magnitude larger than the individual laser wavelengths. The distance between interferometric fringes is now dependent on this beat frequency, which can be adjusted by tuning the wavelengths of the two lasers. The beat wavelength as a function of the wavelengths of the two lasers is given by:
\begin{align}
    \lambda_{MW} &= \frac{\lambda_1 \lambda_2}{|\lambda_1 - \lambda_2|}
\end{align} 
or simply $\nu_2 - \nu_1 =\nu_{MW}$ in terms of frequency. For the b-direction measurement (A - C), the satellite bus provides positioning knowledge at the millimeter scale for missions such as SEIRIOS, well within the synthetic wavelength’s unambiguous range of 1.5 cm.

%%%%%%%%%%%%%%%%%%%%%%%%%%%%%%%%%%%%%%%%%%%%%%%%%%%%%%%%%%%%%%
\section{Experimental Methods}
\label{sec:methods}
The experimental demonstration considers a single mother-daughter pair to verify and test the system performance. For a full implementation, a second set of laser interferometers would be implemented with shared laser and modulator hardware.

Prototyping and subsequent interrogation of the metrology system detailed in this work was conducted in two stages: a single wavelength stage and a multi-wavelength stage. This allowed us to verify DI ranging and phase readout capabilities, after which we incorporated a second wavelength to repeat the same measurements. The two wavelengths interrogate the sensing field simultaneously and independently. The multi-wavelength readout is the differential phase between the two measurements, which is retrieved by computing the phasor difference between the two wavelength measurements. 

\subsection{Optics}
\label{sec:optics}
\begin{figure*}[h!]
    \centering
    \includegraphics[width=0.8\linewidth]{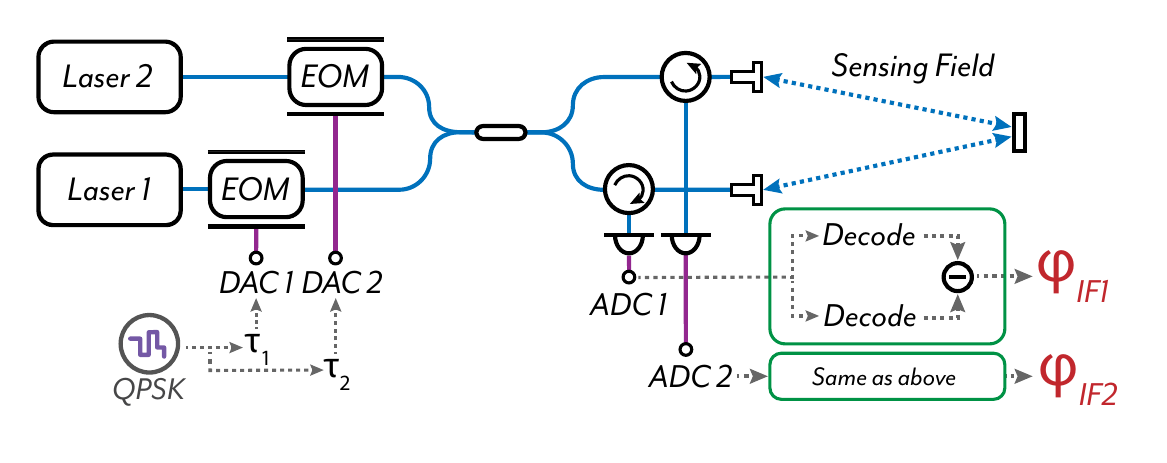}
    \caption{Diagram of the multi-wavelength metrology optics. The single wavelength system is identical, except Laser 2 and its corresponding EOM are inactive. All optics will be housed on the mother satellite excluding the retroreflector to the right of the sensing field, which represents a daughter satellite. (ADC: analog to digital converter, DAC: digital to analog converter, EOM: electro-optic modulator, IF: interferometer)}
    \label{fig:optics}
\end{figure*}
We consider a ranging system comprising two interferometers interrogated by two laser wavelengths. The system uses a RIO Planex 1550\,nm laser diode (Luna Innovations), driven by a Koheron CTL200 digital butterfly laser diode controller. The laser output is fed into an Optilab PM‑1550‑0.2‑S low‑drive electro‑optic phase modulator (EOM), which applies QPSK phase modulation using code generated by a National Instruments NI‑7935R FPGA controller with an NI‑5782 adapter module. Each laser is modulated by its own EOM with a unique QPSK code, enabling identification of the two wavelengths and the corresponding interferometer output. The modulated beams are combined and split into two paths using an AFW 2×2 50/50 PM fiber coupler. We note that in future iterations of the system, additional links could be implemented through further splitting. For this dual interferometer system, each output is routed through an AFW 3‑port PM fiber optic circulator and launched into free space using two Thorlabs F280APC‑1550 fiber collimators, with one mounted a linear motorized T\_LS28 Zaber stage. The beams reflect from a single Thorlabs SM2‑threaded TIR retroreflector, with the separation between the collimators and the retroreflector representing the distance between the mother and daughter satellites. 

\subsection{Digital Signal Processing}
\label{sec:DSP}
\begin{figure*}[h]
    \centering
    \includegraphics[width=0.9\linewidth]{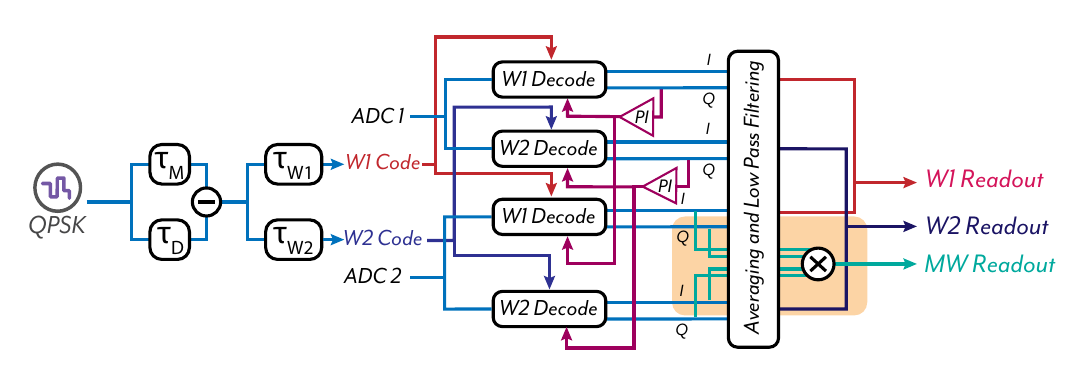}
    \caption{Digital demodulation process. For each wavelength, a local copy of the QPSK code at delays corresponding to the mother and daughter reflections ($\tau_M$ and $\tau_D$) is used to simultaneously decode the returning signals, with the difference between them ($\tau_M-\tau_D$) representing the round trip time of flight between the satellites. The demodulated Q components from one of the interferometers (ADC 1) are used to maintain the interferometer phase at zero (PI controllers).}
    \label{fig:DSP}
\end{figure*}
\begin{figure}[h]
    \centering
    \includegraphics[width=\linewidth]{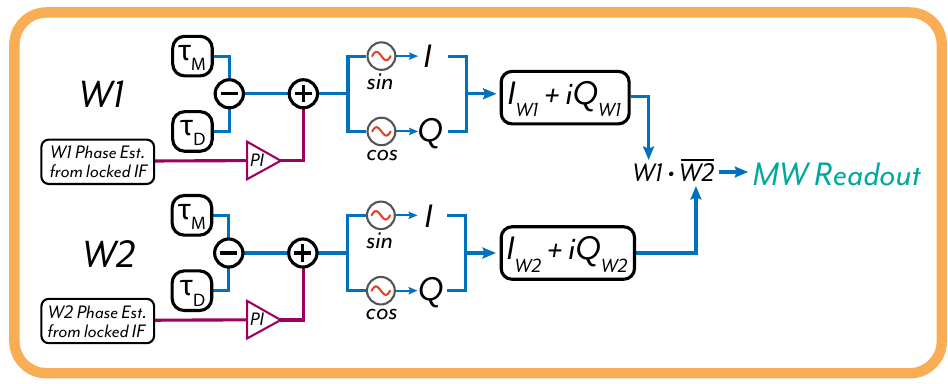}
    \caption{Digital signal processing procedure for the multi-wavelength readout. }
    \label{fig:DSPdetail}
\end{figure}
Each interferometer uses three QPSK codes: one for encoding the outgoing light and two delayed replicas for demodulating the returning signals. The FPGA generates the initial QPSK sequence and drives the EOM, modulating the code onto the outgoing optical field. The returning signal contains two delayed copies of this code: one from the mother reflection (ie. the signal that is internally reflected on the mother satellite) and one from the daughter reflection. To separate these components, the received signal is demodulated twice using the same QPSK sequence but with different time delays for each combination of wavelength, mother delay $\tau_M$, and daughter delay $\tau_D$. The demodulation of these two signals occur simultaneously on the FPGA, allowing independent extraction of the mother and daughter I/Q components. The time of flight, $\tau_D - \tau_M$, is then inferred from the difference between the two demodulation delays for a single wavelength.

In this system, the first interferometer operates as a closed phase‑locked loop system that maintains the interferometer phase at zero by continually updating the demodulation phase using previous measurements. The same demodulation phase signal is fed to the second interferometer, whose output therefore corresponds to the differential phase between the two interferometers.

After demodulation, the I/Q components are low-pass filtered and averaged over a full code period to suppress high‑frequency noise and residual modulation artifacts. The resulting interferometer I and Q values are then transferred to the host computer, where the phase of the complex phasor is computed according to Eqn. \ref{eq:arctan}:
\begin{align}
    \Delta\phi(t) &= \arctan(Q/I)
    \label{eq:arctan}
\end{align}
The recovered phase is used to determine the displacement along the baseline axis, from which the s-direction displacement is derived. The differential path length $L$ between interferometric links $A$ and $B$ is given by:
\begin{align}
    L &= \frac{\Delta\phi(t)\lambda}{2\pi}
    \label{eq:OPD}
\end{align}
where $L = A-B$, and $A$ and $B$ are the round-trip optical path lengths of links $A$ and $B$. 

In a complete system, the absolute separation between the two satellites is provided by a coarse positioning system, such as differential GPS. Using this absolute separation, the displacement along the starlight axis can be determined using Eqn. \ref{eq:sdistance}:
\begin{align}
    s &= \frac{Lb}{2d},
    \label{eq:sdistance}
\end{align}
with $d$ being the separation distance between the two collimators on the mother satellite and $b$ being the baseline distance between the mother and daughter satellites.

%%%%%%%%%%%%%%%%%%%%%%%%%%%%%%%%%%%%%%%%%%%%%%%%%%%%%%%%%%%%%%
\section{Results and Discussion}
\label{sec:results}
\subsection{RMS Noise}
\label{sec:RMS}
\begin{figure*}[h!]
    \centering
    \includegraphics[width=0.9\linewidth]{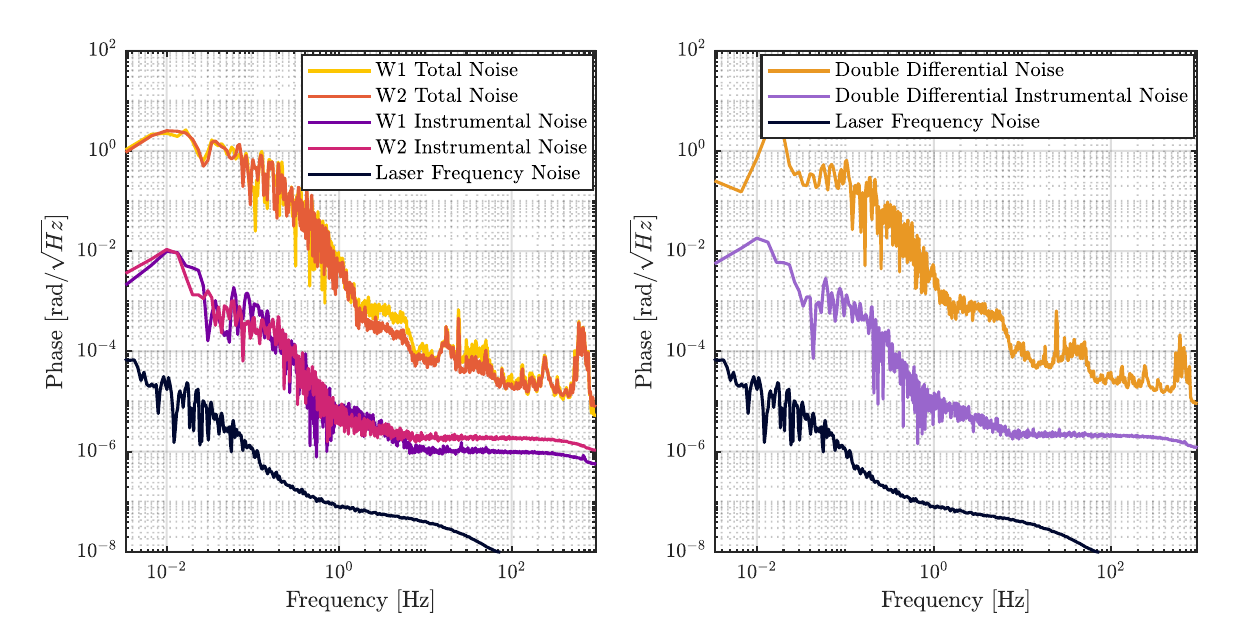}
    \caption{Amplitude spectral densities (ASDs) of the system noise floors. (left) The total noise traces illustrate the direct subtraction measurement for each wavelength under open-path laboratory conditions. The electronic noise traces are a factor of 100 lower. The laser frequency noise is measured via a direct beat note measurement and projected to the phase noise floor. (right) The multi-wavelength total and electronic noise floors computed by taking the difference of the two individual wavelength measurements.}
    \label{fig:PSD}
\end{figure*}
To interrogate the stability of the system over the requisite capture time (10 seconds), we look at the root-mean-squared (RMS) noise associated with both laser wavelengths individually and the synthetic wavelength. 

Our optical noise floor is currently limited by atmospheric turbulence in the beam path. The effect of atmospheric turbulence on the local refractive index is known to cause optical path and signal phase jitter, reducing the stability of time-frequency transfer in the system \citep{liu2022influence}. The electronic noise floor was determined by applying a common photodetector signal to both ADC. This rejects optical noise sources including shot noise, laser intensity noise and polarization fluctuations, as well as environmental fluctuations between interferometers. It therefore sets a lower bound on achievable performance, demonstrating that the ADC front-end and signal demodulation are at the required level. The full optical measurement (Fig. \ref{fig:PSD}) is a factor of 100 higher in phase noise spectral density below 10 Hz, and dominated by air path fluctuations and mechanical noise in the unenclosed beam path.
\begin{figure}[h!]
    \centering
    \includegraphics[width=\linewidth]{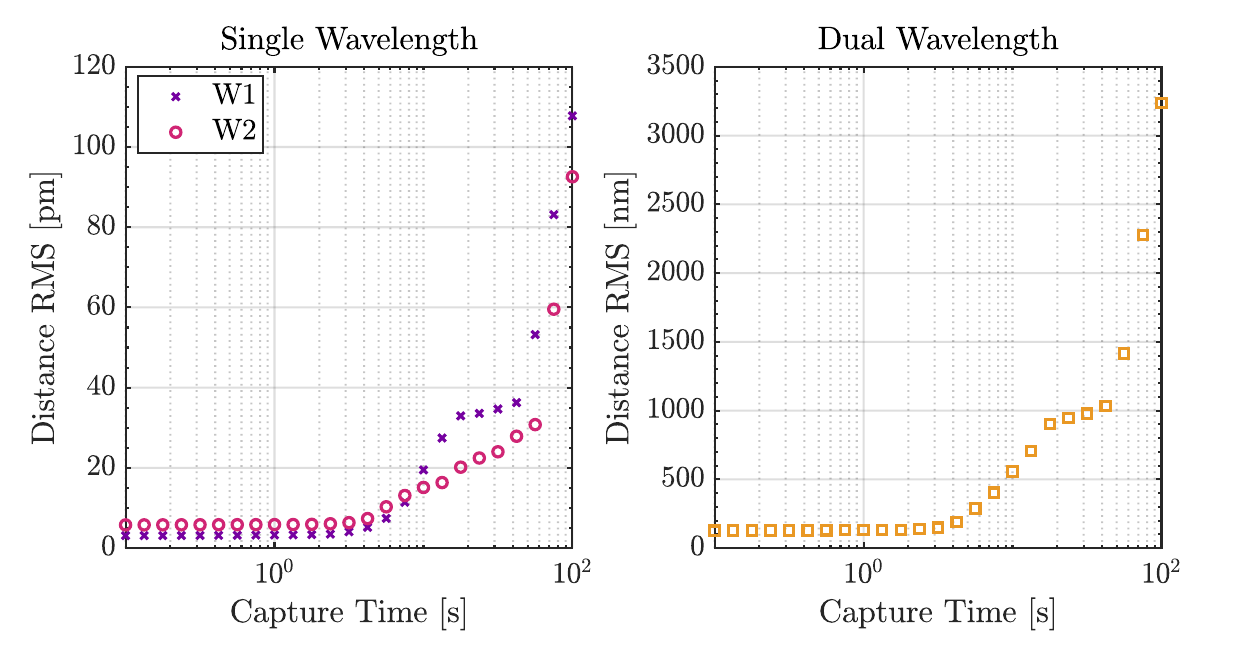}
    \caption{Root mean squares (RMS) for the single wavelength (left) and multi-wavelength (right) electronic noise floors. The RMS values over 10 seconds are 19.48\,pm, 15.11\,pm and 556.16\,nm for W1, W2 and the synthetic wavelength respectively.}
    \label{fig:RMS}
\end{figure}
For the single wavelength measurements (Fig. \ref{fig:RMS}, left subplot), we achieve an RMS of below 20\,pm over a 10 second capture time for both wavelengths, meeting our minimum displacement sensitivity requirement of 560\,pm integrated over 10 seconds. In the case of the beat wavelength measurement (Fig. \ref{fig:RMS}, right subplot), we require that the displacement sensitivity be within the fringe acqusition requirement of 400\,nm. We have achieved an RMS of 556\,nm over a 10 second capture time for the synthetic wavelength measurement, which is currently outside of this requirement. We note that the multi-wavelength measurement calibration is expected to drift, since the two lasers are not frequency stabilized to each other or a common reference. This will be implemented with the next iteration of control electronics for the system. The laser power levels were matched manually using a power meter but are not actively monitored or controlled, which explain why the single wavelength traces to differ from each other. 

\subsection{Laser Frequency Noise Coupling}
\label{sec:LFN}
To understand the laser frequency noise coupling process, we consider a transfer function model for laser frequency noise into our measurement. In the time domain, we consider the phase associated with the interferometer arm length difference for some interferometer $A$ is given by:
\begin{align}
    \phi_A(t) &= \phi_{r_A}(t)- \phi_{r_r}(t),
\end{align}
where $\phi_{r_A}(t)$ is the phase at the $A$ collimator, and $\phi_{r_r}(t)$ is the phase at the retroreflector. We can rewrite this in terms of the input phase $\phi_{in}$ (ie. common phase from the system up until the collimator), and the travel time (delay) between the collimator and the retroreflector $\tau_A$:
\begin{align}
    \phi_A(t) &= \phi_{in}(t)-\phi_{in}(t-\tau_A).
\end{align}
Let the Laplace transform of $\phi_{in}$ be given by $\Phi_{in}(s)=\Laplace\{(\phi_{in}(t)\}$. $\phi_A(t)$ can now be expressed as:
\begin{align}
    \phi_A(t) &= \Laplace\{(\phi_{in}(t)\} - \Laplace\{(\phi_{in}(t-\tau_A)\}\nonumber \\
    &= \Phi_{in}(s) - \Phi_{in}(s)e^{-\tau_As} \nonumber \\
    \Phi_A(s) &= (1-e^{-\tau_As})\Phi_{in}(s)
\end{align}
By extension, a second interferometer $B$ would have a phase $\phi_B(t)$ and a time delay $\tau_B$ associated with its arm length difference, which in the Laplace domain is given by:
\begin{align}
    \Phi_B(s) &= (1-e^{-\tau_Bs})\Phi_{in}(s)
\end{align}
We are interested in the laser frequency noise (LFN) induced due to the path difference (and thus phase difference) between the interferometers $A$ and $B$, noting that the short arm of both interferometers is common. Therefore, we can express the phase difference $\Phi_{AB}(s)$ as:
\begin{align}
    \Phi_{AB}(s) &= \Phi_{A}(s)-\Phi_{B}(s) \nonumber \\
    &= (1-e^{-\tau_As})\Phi_{in}(s) - (1-e^{-\tau_Bs})\Phi_{in}(s) \nonumber\\
    &= \Phi_{in}(s)[e^{-\tau_As}-e^{-\tau_As}]
\end{align}
To obtain the transfer function $H(s)$, we divide this result by frequency $F(s)$. We note that the relationship between frequency and phase is given by $f=\frac{1}{2\pi}\frac{\delta\phi}{\delta t}$. Taking the integral from $s$ to infinity gives us:
\begin{align}
    F(s) &= \frac{s\Phi(s)}{2\pi}
\end{align}
Therefore, the transfer function is:
\begin{align}
    H(s) &= \frac{\Phi_{AB}(s)}{F(s)}\nonumber \\
    &= \frac{\Phi_{in}(s)[e^{-\tau_Bs}-e^{-\tau_As}]}{s\Phi_{in}(s)/2\pi}\nonumber \\
    &= \frac{2\pi[e^{-\tau_Bs}-e^{-\tau_As}]}{s} \text{ rad}/\text{Hz}
\end{align}
The Laplace domain is equivalent to the Fourier domain if all components in $s$ are imaginary, ie. $s=i\omega$, with $\omega=2\pi F(s)$, thus:
\begin{align}
     H(\omega) &= \frac{2\pi[e^{-i\tau_B\omega}-e^{-i\tau_A\omega}]}{i\omega}
\end{align}
Using the relation $\phi = \frac{2\pi}{\lambda} L$, we can rewrite $H(s)$ in units of the length difference between interferometers $A$ and $B$ per Fourier frequency:
\begin{align}
     H(\omega) &= \lambda\frac{[e^{-i\tau_B\omega}-e^{-i\tau_A\omega}]}{i\omega} \text{ m}/\text{Hz}
\end{align}
As we are interested in the path length difference between interferometers $A$ and $B$, we factor out the midpoint of the phase (exponential) terms and simplify the expression:
\begin{align}
    \nonumber H(\omega) &= \frac{\lambda}{i\omega}e^{-i\omega(\frac{\tau_A+\tau_B}{2})}[e^{-i\omega(\frac{\tau_B-\tau_A}{2})}-e^{i\omega(\frac{\tau_B-\tau_A}{2})}] \\
    \nonumber &= \frac{2\lambda}{\omega}e^{-i\omega(\frac{\tau_A+\tau_B}{2})}\sin\left[\frac{(\tau_B-\tau_A)\omega}{2}\right] \\
    &= \lambda(\tau_B-\tau_A)e^{-i\omega(\frac{\tau_A+\tau_B}{2})}\text{sinc}[\frac{(\tau_B-\tau_A)\omega}{2}].
\end{align}
Taking the absolute magnitude cancels the phase terms, yielding:
\begin{align}
    H(\omega) &= \lambda\Delta\tau\;\text{sinc}[\frac{\Delta\tau\omega}{2}],
    \label{eq:transferfunction}
\end{align}
where $\Delta\tau = \tau_B - \tau_A$, $\lambda$ is the central wavelength and $\omega$ is the Fourier frequency of the laser frequency noise. 

The time-of-flight difference for the two interferometers, $\Delta\tau$, is dependent on the maximum s-direction displacement change expected by the system. We take the maximum s-direction displacement to be 5\,cm, based on the worst-case J2 perturbation for satellites in low-Earth orbit. Using Equation \ref{eq:sdistance}, the $\Delta\tau$ associated with an s-displacement of 5\,cm is given by:
\begin{align}
    \nonumber \Delta L &= \frac{2ds}{b} = c\Delta\tau,
\end{align}
where $d$ is the distance between the two interferometer output collimators, and $b$ is the inter-satellite baseline between mother and daughter. Taking the distance between the collimators on the mother to be $d=20$\,cm. We evalulate at the minimum satellite separation, which for a given s-direction tolerance, causes the maximum $\Delta\tau$ between the two sensing paths. Therefore, evaluating at the minimum absolute distance between mother and daughter of $b=5$\,m, we find:
\begin{align}
    \Delta\tau_{\text{max}} &= \frac{2ds}{bc} \\
   \nonumber &= 1.33\times10^{-11} \text{ s}.   
\end{align}
For all Fourier frequencies of interest, $\omega\Delta\tau_{\text{max}} \ll 1$, we can approximate the sinc term as unity. This reduces the transfer function to be independent of $\omega$, yielding $|H| = \lambda\Delta\tau_{\text{max}}$ m/Hz. Multiplying by the wavenumber, $2\pi/\lambda$, gives the equivalent phase transfer function, $2\pi\Delta\tau$ rad/Hz, which we used to convert the experimental LFN trace in Figure \ref{fig:PSD} from frequency to phase.

For $\lambda=1550$\,nm, we can compute the expected worst case length error for a laser frequency error, given by $H = \lambda\Delta\tau_{\text{max}} = 2.07\times10^{-17}$ m/Hz. Our maximum distance error requirement of 560 pm can then be used to determine the maximum $\Delta\nu$ permissible for the system:
\begin{align}
    \nonumber {\Delta\nu_{SW}} &= \frac{560\times10^{-12}\,\text{m}}{2.07\times10^{-17}\,\text{m/Hz}} \\
    \nonumber &= 27 \,\text{MHz}.
\end{align}
For the multi-wavelength system, where $\lambda_{MW} = c / 10\,\text{GHz} =0.03$ m, the transfer function can be similarly computed to be  $H_{MW}(\nu)= 4\times10^{-13}\,\text{m/Hz}$. From the absolute path length requirement of 400 nm, we can compute a maximum $\Delta\nu_{MW}$ of:
\begin{align}
    \nonumber \Delta\nu_{MW} &= \frac{400\times10^{-9}\,\text{m}}{4\times10^{-13}\text{ m/Hz}} \\
    \nonumber &= 1\,\text{MHz}.
\end{align}
Therefore, frequency noise induced phase noise sits below the sensitivity requirements as long as the individual laser drift is below 27 MHz, and the frequency separation between them is controlled to within 1 MHz. With our current hardware, we are able to meet the requirements for the single wavelength readout. Further stabilization is required, by offset sideband locking, in order to meet requirements for the wavelength separation. 

\subsection{Micrometer Stage Measurements}
\label{sec:micrometer}
\begin{figure}[h]
    \centering
    \includegraphics[width=\linewidth]{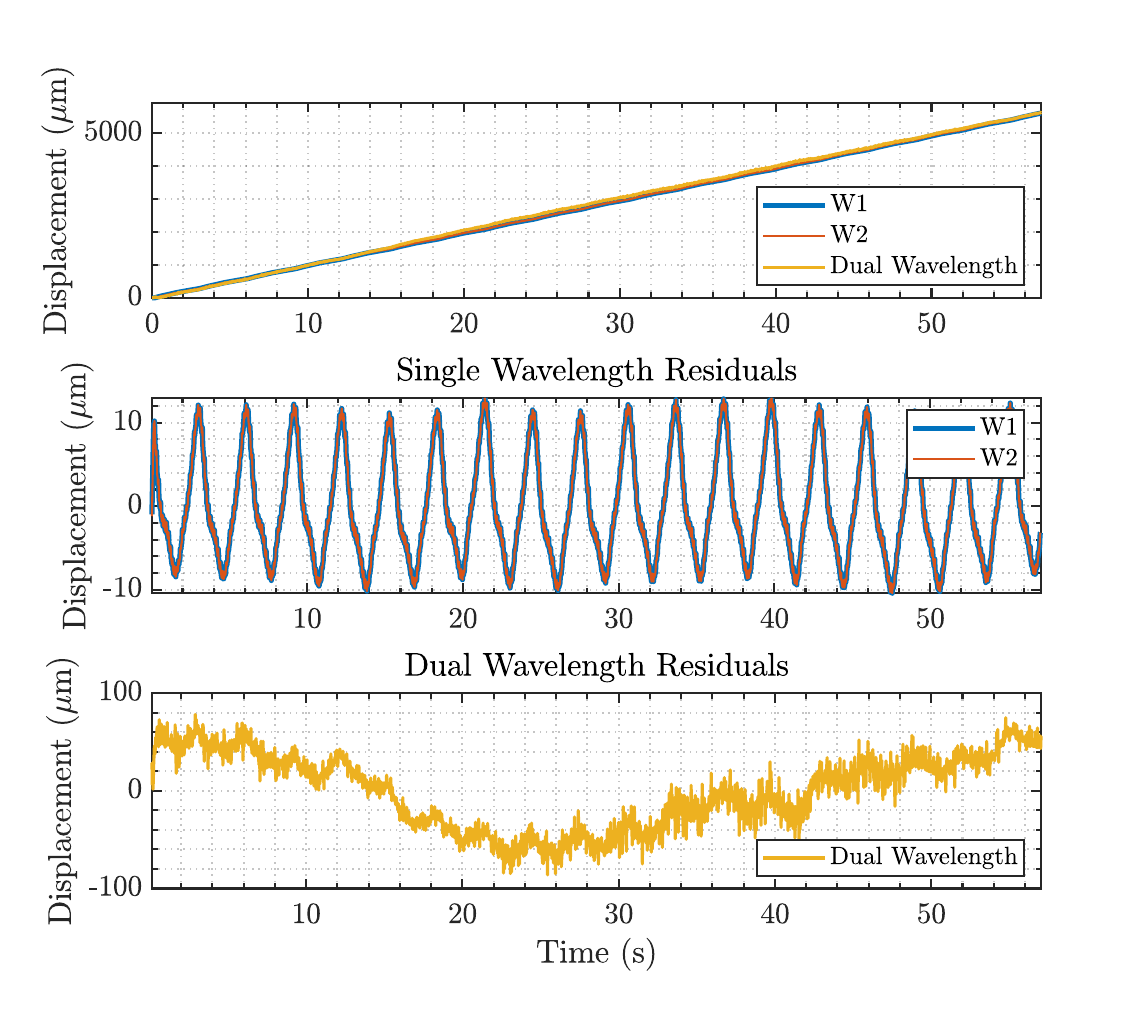}
    \caption{The two single wavelength measurements and dual-wavelength measurement of the 100\,\textmu m/s sweep is shown in the top subplot. Residuals for the single wavelengths and dual-wavelength sweep are shown in the middle and bottom subplots respectively.}
    \label{fig:zaber}
\end{figure}

To characterize the displacement and velocity sensitivity, we dynamically translate one collimator along its beam axis with a linear motorized T\_LS28 Zaber stage. This changes the length of one of the interferometric links at a constant velocity of 100\,$\mu$m/s, which is equivalent to a change in length produced by s-direction motion of the daughter satellite as per Eqn. \ref{eq:sdistance}. The results of this measurement are shown in Figure \ref{fig:zaber}, where the residuals are the measured displacement minus the expected displacement assuming a velocity of 100\,$\mu$m/s. 
We can quantify the linearity of the system by examining the residual of the displacement readouts over time, since the measurement sweep itself was at a constant velocity. This is done by comparing the deviation of the measured displacements from the target velocity with the total displacement over the entire sweep. 

For the single wavelength residuals, we see a periodic displacement error of roughly $\pm 10$\, $\mu$m. This can be attributed to the Zaber stage, namely linear motion per motor revolution (reported as 0.6096\,mm on its specification sheet). At a speed of 100\,$\mu$m/s, a displacement of 0.6096\,mm occurs approximately every 6 seconds. We see that the displacement residuals peak every 3 seconds, a periodic error that we attribute to the lead screw driving the linear stage motion.

On the other hand, the error in the multi-wavelength measurement is likely to have been caused by cyclic error. The high frequency dynamics we see in the multi-wavelength residual we attribute to modulation induced cyclic errors in the individual optical readout. These errors are well-known for DI and in the multi-wavelength readout, the effect is amplified by the ratio of the synthetic wavelength to the optical wavelength. We are currently investigating methods to reduce these errors either by upconversion or through feedforward cyclic error correction \citep{Hu2015,sambridge2025dual-quadrature-b90}. Secondly, there may be optical and/or electronic cross-talk (signal mixing) between the wavelength channels. This would be impacted by the bandwidth limits of the RF electronic system available to the experiment. This error would have a period of the synthetic wavelength, and may be the cause of the large drift motion observed in the scan. A larger scan range on the stage is required to confirm this for the particular multi-wavelength frequency used. The exact source of the cyclic error and identifying the appropriate method to mitigate it is currently being investigated and will be reported as part of future work.

%%%%%%%%%%%%%%%%%%%%%%%%%%%%%%%%%%%%%%%%%%%%%%%%%%%%%%%%%%%%%%
\section{Conclusion}
\label{sec:conclusion}
In this work, we have presented a laboratory prototype of an inter-satellite metrology system for formation-flying interferometry. To achieve this, three metrology techniques are employed: Digital Interferometric ranging, multi-wavelength interferometry, and digitally enhanced homodyne interferometry. We performed static and dynamic measurements for both single wavelength and multi-wavelength readouts. 

We measure an electronic RMS noise below 20 pm over 10 second integration times for each individual wavelength, below the 560 pm requirement. For the multi-wavelength readout, the electronic noise floor is 556 nm at 10 seconds, with a minimum below 3 seconds integration time of approximately 200 nm. During a dynamic measurement with a 100 $\mu$m/s sweep, the multi-wavelength readout shows a slow varying drift on the order of 100 $\mu$m, which we attribute the cyclic errors in the readout, and will be the focus of future work.

%%%%%%%%%%%%%%%%%%%%%%%%%%%%%%%%%%%%%%%%%%%%%%%%%%%%%%%%%%%%%%
\begin{acknowledgements}
AOC is supported by an Australian Government Research Training Program (RTP) Scholarship. CPB is an Associate Investigator with the Australian Research Council OzGrav and COMBS Centres of Excellence. This research was supported by the Australian Government through the Australian Research Council’s Discovery Projects funding scheme (project DP240102015). 
\end{acknowledgements}
% Please note that we have included the references below in
% order to compile the document, but we ask you to:

\bibliographystyle{bibtex/aa.bst} % style aa.bst
\bibliography{references.bib} % your references Yourfile.bib

\end{document}